\documentclass[sn-mathphys,Numbered,pdflatex]{sn-jnl}

\usepackage{graphicx}%
\usepackage{multirow}%
\usepackage{amsmath,amssymb,amsfonts}%
\usepackage{amsthm}%
\usepackage{mathrsfs}%
\usepackage[title]{appendix}%
\usepackage{xcolor}%
\usepackage{textcomp}%
\usepackage{manyfoot}%
\usepackage{booktabs}%
\usepackage{algorithm} 
\usepackage[algo2e]{algorithm2e}
\usepackage{algpseudocode}%
\usepackage{listings}%

\begin{document}

\title[Article Title]{An RNN-policy gradient approach for quantum architecture search}


\author[1]{\fnm{Gang Wang} \sur{}}

\author*[1]{\fnm{Bang-Hai Wang} \sur{}}\email{bhwang@gdut.edu.cn}
\equalcont{These authors contributed equally to this work.}
\author[2,3]{\fnm{Shao-Ming Fei} \sur{}}
\equalcont{These authors contributed equally to this work.}


\affil*[1]{\orgdiv{School of Computer Science and Technology}, \orgname{Guangdong University of Technology}, \city{GuangZhou}, \postcode{510006}, \state{GuangDong}, \country{China}}
\affil[2]{\orgdiv{School of Mathematical Sciences}, \orgname{Capital Normal University}, \city{Beijing}, \postcode{100048}, \country{China}}
\affil[3]{\orgdiv{Max-Planck-Institute for Mathematics in the Sciences}, \postcode{04103}, \city{Leipzig}, \country{Germany}}



\abstract{Variational quantum circuits are one of the promising ways to exploit the advantages of quantum computing in 
the noisy intermediate-scale quantum technology era. The design of the quantum circuit architecture might greatly affect 
the performance capability of the quantum algorithms. The quantum architecture search is the process of automatically 
designing quantum circuit architecture, aiming at finding the optimal quantum circuit composition architecture by the 
algorithm for a given task, so that the algorithm can learn to design the circuit architecture. Compared to manual design, 
quantum architecture search algorithms are more effective in finding quantum circuits with better performance capabilities. 
In this paper, based on the deep reinforcement learning, we propose an approach for quantum circuit architecture search. 
The sampling of the circuit architecture is learnt through reinforcement learning based controller. Layer-based search is 
also used to accelerate the computational efficiency of the search algorithm. Applying to data classification tasks we 
show that the method can search for quantum circuit architectures with  better accuracies. Moreover, the circuit has a 
smaller number of quantum gates and parameters.}




\maketitle

\section{Introduction}\label{sec1}

The development of quantum computing and quantum information theory has revolutionized the implementation of computing.
Quantum computing can theoretically break the RSA algorithm in a few hours \cite{shor1999polynomial}, which would take 
billions of years in a traditional computational environment. Meanwhile, machine learning has achieved great success in 
the last two decades, with major breakthroughs in applications such as image processing \cite{lecun1998gradient}, language 
processing \cite{sutskever2014sequence} and games  \cite{silver2016mastering}. Integrating quantum computing and machine 
learning has become a popular research topic. On the one hand, machine learning methods are used to solve quantum physics 
problems such as quantum many-body states \cite{carleo2017solving}, nonlocality detection \cite{deng2018machine}, quantum 
error correction \cite{baireuther2018machine,andreasson2019quantum} and quantum control \cite{an2019deep}. On the other 
hand, quantum computing is expected to bring new developments to traditional machine learning, e.g., quantum support vector 
machines \cite{rebentrost2014quantum,li2015experimental}, quantum reinforcement learning\cite{jerbi2021parametrized}, etc.

One of the most promising ways to exploit the advantages of quantum computing on the noisy intermediate-scale quantum devices 
\cite{preskill2018quantum} is currently the hybrid quantum classical models which use classical computing methods to 
optimize the parameters of the parameterized quantum circuits, called variational quantum circuits \cite{bharti2022noisy}. 
This kind of hybrid quantum models includes quantum neural networks \cite{beer2020training} and quantum eigen-solvers 
\cite{peruzzo2014variational} etc.. In recent years, typical applications are given to quantum eigen-solvers for ground 
state estimation of quantum systems \cite{liu2019variational}, quantum approximate optimization algorithms for better 
approximation of NP-hard combination optimization problems \cite{farhi2014quantum} and solving machine learning tasks in 
classical settings by quantum machine learning \cite{farhi2018classification}. Although good performances have been 
experimentally verified in tasks such as small-scale energy estimation \cite{havlivcek2019supervised} and data classification 
\cite{huang2021experimental,hur2022quantum}, it is still a big problem to design the architecture of quantum circuits. As 
the number of quantum gates and the depth of the circuit increase, the noise of the circuit also increases. As a result, 
this leads to the barren plateau \cite{mcclean2018barren} and affects the performance of the variational circuit. Thus, 
efficient designs to discover the optimal representation for quantum circuit architectures with restricted scale are 
critical research challenges.

Similarly, the design of neural networks in deep learning depends significantly on the researchers' previous knowledge and 
experience, which may limit the design of neural network architectures. Therefore, neural architecture search (NAS) 
\cite{ren2021comprehensive} has been proposed aimed to design neural network structures automatically with restricted 
computational resources and the least manual intervention, and to achieve optimal performance as good as possible. Typical 
approaches include reinforcement learning \cite{zoph2016neural,pham2018efficient,cui2019fast}, evaluation algorithms 
\cite{real2017large} and differentiable search \cite{liu2018darts}. The problems and objectives faced by neural architecture 
search and quantum architecture search are rather similar. Inspired by neural architecture search, the related quantum 
circuit architecture search based on reinforcement learning \cite{ostaszewski2021reinforcement}, evaluation algorithm 
\cite{lu2021markovian} and differentiable search \cite{du2022quantum,zhang2020differentiable} have been proposed as well. 
In contrast, related algorithms based on reinforcement learning are currently mostly used for quantum control \cite{an2019deep} 
and quantum compiling \cite{zhang2020topological,gong2021weighted}. Architecture search has not been directly applied for a 
given computational task.

In this paper, we propose a reinforcement learning based quantum circuit architecture search method for searching optimal 
ansatz in quantum circuits. We use a recurrent neural network(RNN) to fit the policy function and optimize it with Monte 
Carlo policy gradient\cite{williams1992simple} as the strategy for architecture search. The computational efficiency of the 
search is also improved by using layer-based search. To verify the efficiency of the method, we use the MNIST dataset 
\cite{MNIST} for classification. It outperforms manual-based designs and employs fewer quantum gates and variable parameters.

\section{PRELIMINARY}

\subsection{Neural architecture search}

NAS is an emerging research area in AutoML aimed to automatically find the optimal performance neural network architecture 
for a given task. The use of reinforcement learning to search for neural network architecture in image classification beyond 
the previously manual design of neural networks in 2017 \cite{zoph2016neural} is a significant event in the field of NAS. 
The original NAS approach requires a large amount of computational resources. A more efficient neural architecture search 
(ENAS) was subsequently proposed to address the problems for which previous NAS methods require large computational resources 
\cite{pham2018efficient}. This method denotes the search space as a directed acyclic graph, where any subgraph represents a 
network architecture, replacing the global search space with a cell-based search space. At the same time, different network 
architectures share the parameters on the whole directed acyclic graph nodes, i.e., weight sharing. ENAS reduces significantly 
the expense and improves the computational efficiency compared to the previous NAS. The modular search space and weight 
sharing effectively reduce the expenses of neural architecture search. Later FPNAS \cite{cui2019fast} also considered the 
diversity of the cell stacking, and demonstrated that stacking different cells is effective in improving the performance of 
the neural architecture. 
In addition to reinforcement learning-based strategies, there are also search methods based on evaluation algorithms 
\cite{real2017large}, Bayesian optimization \cite{kandasamy2018neural}, SMBO (Sequential Model-Based Global Optimization) 
\cite{liu2018progressive} and Monte Carlo tree search \cite{negrinho2017deeparchitect}. These methods treat neural 
architecture search as a black-box optimization problem in discrete search strategies. By converting discrete spaces into 
continuous differentiable formats, the DARTS method was proposed in \cite{shin2018differentiable}. As with ENAS, the network 
space is represented as a directed acyclic graph. The node connections and activation functions are combined into a matrix, 
where each element represents the weight of the connection and activation function, and the Softmax function is used in the 
search to convert the search space into a continuous space and the objective function into a differentiable function. The 
DARTS search neural network architecture performs well under classical computational tasks and enhances computational 
efficiency compared to previous reinforcement learning approaches. Subsequently, many variants based on DARST were 
proposed\cite{liang2019darts+,chu2020darts}. With the success of the attention mechanism and Transformer in the field of 
natural language processing, researchers use the attention mechanism in other application areas. A Transformer-based 
approach to encoding computational awareness architecture was used in \cite{yan2021cate}, which results in neural network 
architectures with dense computational resource consumption in the hidden space. The results indicate that this encoding 
approach can enhance downscale tasks.

\subsection{Quantum architecture search}

Inspired by the research on neural architecture search, researchers have also proposed a series of algorithms for quantum 
circuit architecture search. For the quantum circuit in image classification, Hur et al. \cite{hur2022quantum} considered 
the effect of different QCNN's circuit ansatz architecture on the classification performance, through different structures 
are still designed manually. Du et al. \cite{du2022quantum} used an approach of differentiable search for quantum 
circuit architecture with decent resources and runtime efficiency, in addition to excellent performance on classification 
tasks and hydrogen molecule ground state estimation tasks. Similarly, Zhang et al. \cite{zhang2020differentiable} 
proposed the differentiable quantum architecture search (DQAS) that was experimentally demonstrated to be superior to the 
traditional QAOA circuit on the MaxCut problem. Ostaszewski et al. \cite{ostaszewski2021reinforcement} proposed a 
reinforcement learning algorithm that can automatically explore the space of possible architectures and recognize economic 
loops that still generate accurate ground state energy estimation. The algorithm uses a feedback-driven learning approach 
that adapts the complexity of the learning problem to the current performance of the learning algorithm and progressively 
improves the accuracy of the results while reducing the circuit depth to a minimum at the same time. Lu et al. 
\cite{lu2021markovian} proposed the use of neuroevolutionary algorithms for quantum circuit architecture search for different 
machine learning tasks, establishing a one-to-one mapping between quantum circuits and directed graphs, and reducing the 
problem of searching for proper gate sequences to the task of searching for proper paths in the corresponding graphs as a 
Markov process. 

In addition to the applied circuit architecture search in problems of quantum machine learning, similar circuit search methods 
can be applied in quantum compilers as well to generate specific quantum gate sequences. Kuo et al. \cite{kuo2021quantum} 
used A2C (Advantage actor-critic) and PPO (Proximal policy optimization) algorithms for quantum architecture search to 
demonstrate the successful generation of quantum gate sequences with multi-GHZ states without encoding any knowledge of 
quantum physics in the agent. Zhang et al. \cite{zhang2020topological} propose a deep reinforcement learning-based 
algorithm which is designed to compile an arbitrary single quantum bit gate from a finite common set into a sequence of 
fundamental gates, which is applied to the topological compilation of Fibonacci arbitrators to obtain an approximate optimal 
sequence of arbitrary single qubits of unitary. Subsequently, PPO is further used to compile arbitrary quantum channels\cite{gong2021weighted}. 
The current research for the field of quantum artificial intelligence including quantum circuit architecture search is 
still far from being satisfied. How to design or optimize quantum algorithms to solve quantum computing problems is an 
important challenging problem. The extension of classical computing algorithms is also in the preliminary stage. Quantum 
architecture search using reinforcement learning is currently mostly evolved in quantum compilation. This paper aims to 
investigate a reinforcement learning quantum circuit architecture design based approach for a given task.

\subsection{Variational quantum circuits}

Variational quantum circuits (or parametric quantum circuits) are based on hybrid quantum and classical computational methods. 
A set of quantum gates with parameters $\theta_i$ are carried out on a quantum computer, while the $\theta_i$ is optimized on a 
classical computer, as shown in Fig. 1. {Where the parameters $\theta_i$ are different for each quantum gate.} Usually it includes three steps: encoding, variational and readout. To be processed 
on quantum devices, classical data need to be encoded into quantum states. The encoding layer encodes $n$ dimensional data 
onto $m$ qubits by such as direct encoding, angular encoding, amplitude encoding etc.. The direct encoding represents the 
classical data in binary form and directly maps them onto quantum computational bases. This method takes $n$ qubits to load 
the binary representation of a classical data point $x=(x_1, x_2, ..., x_n)$ and encodes it as $|\Psi_x\rangle=\bigotimes^n_i|x_i\rangle$. 
The angular encoding loads the classical data as the radians of rotation gates acting on qubits. {A classical data 
$x=(x_1, x_2, ..., x_n)$ is encoded as $|\Psi_x\rangle=\bigotimes^n_i cos(x_i)|0\rangle + sin(x_i)|1\rangle$.} The encoding method used in this paper is the 
amplitude encoding, which embeds classical data into the amplitudes of a quantum state. An $n$-dimensional normalized 
classical vector $x$ is encoded to the amplitudes of quantum state by $|\Psi_x\rangle=\sum^n_{i=1}x_i|i\rangle$, 
where $|i\rangle$ is the $i$-th computational basis state and $\sum|x_i|^2=1$

The step variational layer operates on qubits by rotating parametric quantum gates denoted by $U(\theta)=\prod^L_{l=1}U_l(\theta)$, 
where $U_l(\theta)$ consists of a sequence of parameterized quantum gates including single-qubit and two-qubit quantum gates, {$\theta = \{\theta_1, \theta_2, ..., \theta_n\}$ denotes the parameters of all quantum gates in this layer.} 
The performance of the circuit can be improved by increasing the number of variational layers and quantum gates.

After the variational operation of the $L$ layer, the quantum data are read out by measurement. The measurement output is 
$\langle\Psi_x|U(\theta)MU^\dagger(\theta)|\Psi_x\rangle$, where $|\Psi_x\rangle$ is the input quantum state and $M$ are 
measurement operators on read out qubits. Then, the classical computer calculates the loss function $J$ to optimize 
the circuit parameter $\theta$. In this way, the circuit can fit arbitrary functions like traditional neural networks\cite{oh2020tutorial}. Hence, 
QVC is also called quantum neural network and is widely used in quantum machine learning. A series of derivative structures 
such as QCNN \cite{cong2019quantum} and QRNN \cite{bausch2020recurrent} have been proposed.

\begin{figure}[H]
\centering
\includegraphics[width=1\columnwidth]{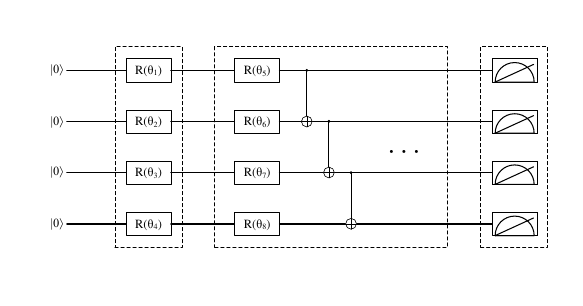}
\caption{The general form of variational quantum circuits}
\label{Fig.1}
\end{figure}

\subsection{Reinforcement learning}

Reinforcement learning emphasizes that the intelligence learns by interacting with the environment and does not require 
imitable supervised signals or complete modeling of the environment. During the continuous interaction between the agent 
and the environment, the agent selects an action, while the environment responds to this action and moves to a new state. 
At the same time, the environment generates a reward. During the interaction the agent needs to maximize the cumulative 
rewards obtained. Reinforcement learning can be modeled as a partially observable Markov decision process defined as a 
tuple $(S, A, T, R, \Omega, O)$, where $S$ is the state set, $A$ is the action set, $T$ is the transfer function that maps 
the state-action to the probability distribution of the next state, $R: S\times A \to R$ is the reward function, $O$ is 
the observation set and $\Omega$ is the probability distribution function that maps observations to states. The agent 
obtains observations $o$ as partial information about the environment state $s$. After the agent selects an action, the 
environment transfers to the next state $s_1$ according to $T$, while giving back new observations $o$ and reward $r$ to 
the agent. The agent's goal is to obtain the maximum cumulative expected reward during the interaction, defined as 
$G=\sum_{t=0}^T\gamma^t R_t$, where $\gamma$ is the discount factor. In this paper, we model quantum architecture search as 
a process of reinforcement learning, as shown in Fig. 2. The state is the current circuit architecture. The action is the 
next layer that will be added to the circuit. The reward is the performance of the current circuit.

\begin{figure}[H]
\centering
\includegraphics[width=1\linewidth]{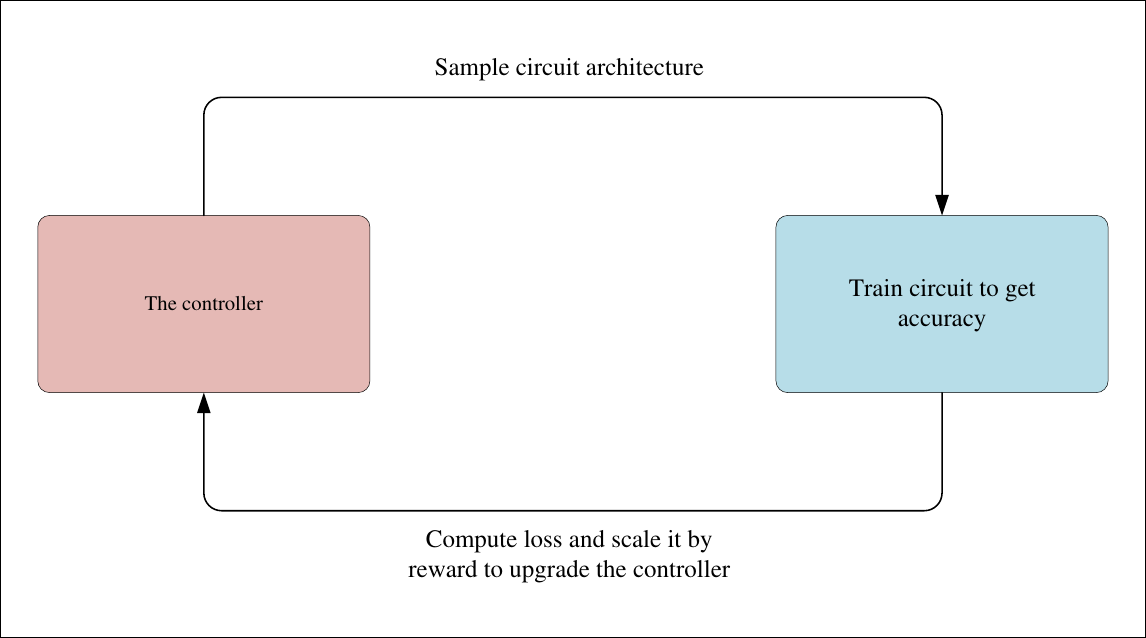}
\caption{Workflow}
\label{Fig.3}
\end{figure}

\section{Method}

The method consists of two parts training process. According to a given task and the existing circuit architecture, a 
controller samples the quantum circuit architecture of each layer based on reinforcement learning and recurrent neural 
network(RNN). The training controller is aimed to find the required gates for the quantum circuit for the given task. The 
next is the process of training circuit. The circuit is composed based on the architecture of the single layer sampled from 
the controller for the given task. The training objective is to optimize the parameters of the circuit. At the end of one 
epoch of training for the circuit, the accuracy performance of the circuit on the validation set is feedback to the 
controller as a reward signal for parameter optimization of the controller network. The training of the circuit and the 
training of the controller constitute a complete epoch of training. We show the details below. 
\subsection{Training process}

The training process shown in Fig. 2 consists of two parts. At the beginning of an epoch, the quantum circuit is initialized to empty. 
The controller selects a quantum gate sequence of a layer to join the circuit based on the initial state of the circuit. The quantum 
circuit adds that sequence to the existing architecture to construct the quantum circuit. Subsequently, Adam is 
used to optimize the parameters of the circuit. The mean
squared error loss is chosen as the loss function $J$ for the training circuit,
\begin{equation}
J=\frac{1}{N} \sum_{i=1}^N\left(y_i-y_{\text {prediction }}\right)^2.
\end{equation}
Where $y_i$ is the target label of input, $y_{prediction}$ is the circuit prediction. The training algorithm is shown in Algorithm 1.
\begin{algorithm}[H]
\caption{Train Circuit}\label{algorithm1}
\While{$epoch < max\ epoch$}{
Sample a random batch $B$ from train set\\
Input circuit to get prediction $y_{prediction}$\\
Compute loss by MSE:\\
$J=\frac{1}{N}\sum^N_{i=1}(y_i-y_{prediction})^2$\\
Upgrade parameters:\\
{
$m_t = \beta_1 m_{t-1} + (1-\beta_1)\nabla_\theta J$\\
$n_t = \beta_2 n_{t-1} + (1-\beta_2)\nabla_\theta J$\\
$\hat{m}_t = \frac{m_t}{1-\beta_1^t}$\\
$\hat{n}_t = \frac{n_t}{1-\beta_2^t}$\\
$\theta_t = \theta_{t-1} - \frac{\alpha}{\sqrt{\hat{n_t} + \varepsilon}} \hat{m_t}$\\
}
}
\end{algorithm}

After the training of the quantum circuit is finished, we compute the accuracy on the validation set as a reward in order to have better 
generality of our controller. The controller parameters are then optimized, with the training of the controller given in the next section. 
Finally, the performance of the circuit representation is verified on the test set. The complete training procedure of our method is shown 
in Algorithm 2.
\begin{algorithm}[H]
\caption{Train procedure}\label{algorithm2}
initialized arch to empty\\
\For{
$epoch<max\ epoch$}{
\For{$layer<max\ layer$}{
Sample a layer L architecture by the controller\\
Add L to arch\\
Generate circuit by arch\\
Train circuit\\
Calculate reward by circuit train loss\\
Upgrade the controller parameters\\
}}
\end{algorithm}



\begin{figure}[]
\centering
\includegraphics[width=1.0\textwidth]{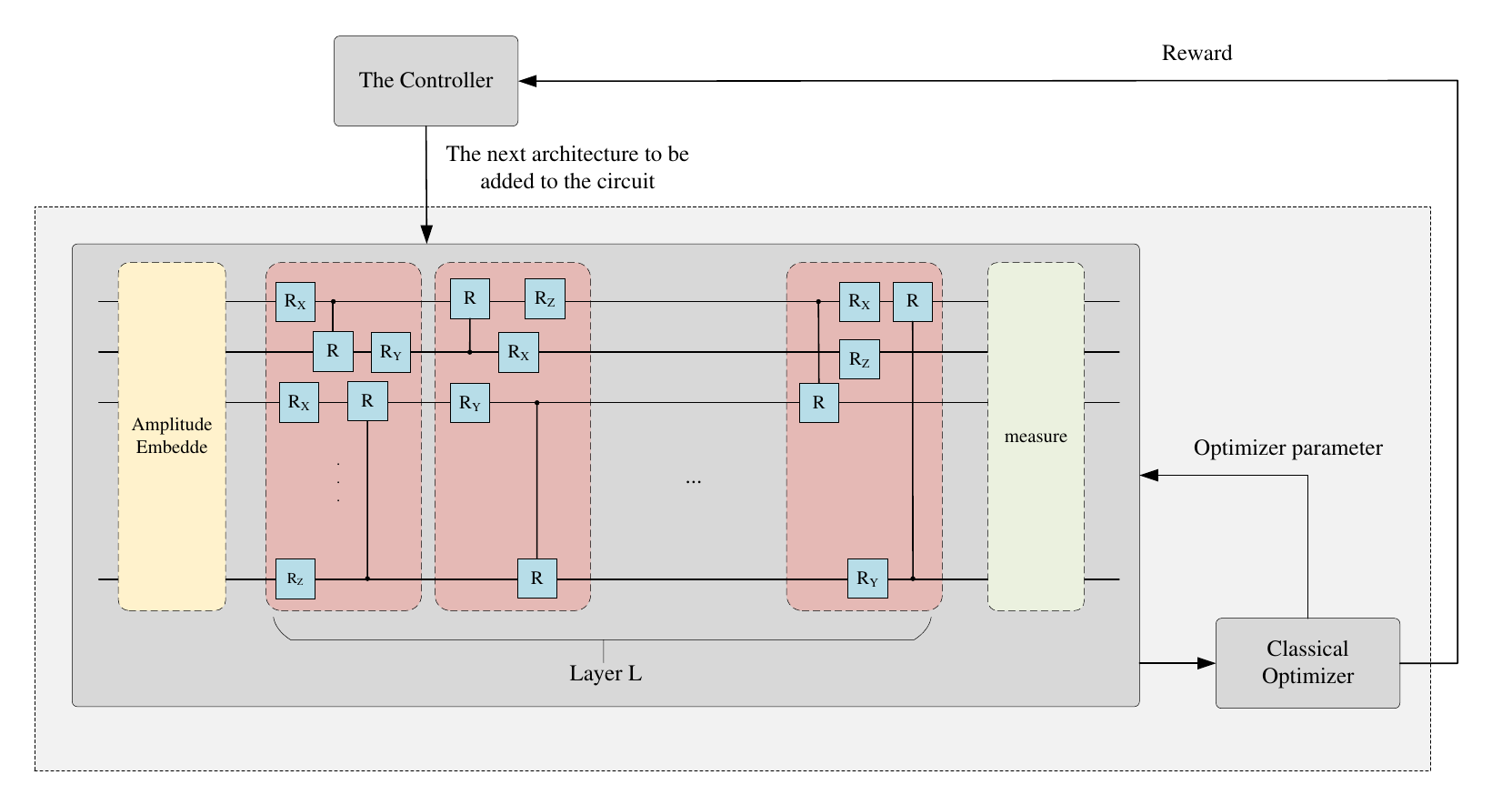}
\caption{Training process}
\label{Fig.4}
\end{figure}

The controller uses a policy gradient approach to optimize the parameters. To find the optimal architecture, the accuracy performance of 
the circuit on the validation set is feedback to the controller as a reward. The goal of controller learning is to maximize the total 
expected reward, i.e:
\begin{equation}
\max J\left(\theta_c\right)=\max E_{P\left(a_t, \theta_c\right)}[R],
\end{equation}
where $a_t$ denotes the circuit architecture at time $t$, $\theta_c$ is the circuit parameter, and $R$ denotes the reward calculated based on the performance of the circuit on the validation set. For optimization, we use the Monte Carlo policy gradient \cite{williams1992simple}\:
\begin{equation}
\nabla_{\theta_c} J\left(\theta_c\right)=\sum_{t=1}^T E_{P\left(a_t, \theta_c\right)}\left[\nabla_{\theta_c} \log P\left(a_t \mid a_{t-1}, \theta_c\right) R\right].
\end{equation}

The complete training process is shown in the Fig. 4. At the beginning of an epoch, the controller samples a random batch from valid set 
of input circuits to compute the loss. Then reward is computed by loss and feeded back to the controller. At last, the controller upgrades 
the policy function. The complete algorithm for controller training is shown in Algorithm 3.
\begin{algorithm}[H]
\caption{Train Controller}\label{algorithm3}
Sample a random batch $B$ from valid set\\
Input circuit to compute loss $L$\\
calculate reward $r$ by $L$\\
calculate return:\\
$R=\sum_{t=0}^T\gamma^tr_t\\
J(\theta_c)=E_{(a_t,\theta_c)}[R]$\\
calculate gradient:\\
$\nabla_{\theta_c} J\left(\theta_c\right)=\sum_{t=1}^T E_{\left(a_t, \theta_c\right)}\left[\nabla_{\theta_c} \log P\left(a_t \mid a_{t-1}, \theta_c\right) R\right]$\\
Upgrade parameters:\\
{
$m_t = \beta_1 m_{t-1} + (1-\beta_1)\nabla_\theta J$\\
$n_t = \beta_2 n_{t-1} + (1-\beta_2)\nabla_\theta J$\\
$\hat{m}_t = \frac{m_t}{1-\beta_1^t}$\\
$\hat{n}_t = \frac{n_t}{1-\beta_2^t}$\\
$\theta_t = \theta_{t-1} - \frac{\alpha}{\sqrt{\hat{n_t} + \varepsilon}} \hat{m_t}$\\
}
\end{algorithm}

\subsection{Controller}
{
We use a reinforcement learning algorithm for policy gradients to design a controller for searching circuit architecture. 
First, the circuit architecture search process is modelled as a Markov decision process.The observation $O_t$ acquired by 
the controller at moment $t$ is the structure $arch_t$ of the quantum variational circuit at that moment in time. $arch_t$ 
consists of the quantum gates in a given set of quantum gates $R$. The quantum gates in the set of available quantum gates 
$R = \{R_x, R_y , R_z , CNOT\}$, where $R_x, R_y , R_z$ are rotational gates and the $CNOT$ is the controlled-NOT operator, denoted as.
\begin{equation}
R_x(\phi)=\left[\begin{array}{cc}
\cos (\phi / 2) & -i \sin (\phi / 2) \\
-i \sin (\phi / 2) & \cos (\phi / 2)
\end{array}\right],
\end{equation}
\begin{equation}
R_y(\phi)=\left[\begin{array}{cc}
\cos (\phi / 2) & -\sin (\phi / 2) \\
\sin (\phi / 2) & \cos (\phi / 2)
\end{array}\right],
\end{equation}
\begin{equation}
R_z(\phi)=\left[\begin{array}{cc}
e^{-i \phi / 2} & 0 \\
0 & e^{i \phi / 2}
\end{array}\right],
\end{equation}
\begin{equation}
CNOT=\left[\begin{array}{cccc}
1 & 0 & 0 & 0 \\
0 & 1 & 0 & 0 \\
0 & 0 & 0 & 1 \\
0 & 0 & 1 & 0
\end{array}\right].
\end{equation}
The controller chooses the action $a_{t+1}$, i.e., the next moment as a layer to join the quantum circuit of $l$ quantum gates 
according to the policy $\pi$, denoted as $a_{t+1} = \pi(arch_t)$. For each quantum gate $G$ in the circuit can be denoted by $\{B, E, T\}$, 
where $B$($E$) is the first (second) qubit in a double quantum operation, and $T$ denotes the type of the quantum gate $T \in R$. When $B$ 
is equal to $E$, the quantum gate is a single quantum operation. There are $4q + q(q-1)$ quantum gates of different types in the 
circuit, where $q$ is the number of qubits in the circuit. Essentially, $4q$ represents the number of single-qubit gates, 
and $q*(q - 1)$ refers to the number of CNOT gates utilized. Each quantum gate is numbered using a unique index in that space to 
facilitate input to the neural network for processing. Each time $l$ quantum gates are added to the circuit, their parameters are 
optimized according to the given task. The controller network is optimized based on the circuit's accuracy performance on the 
validation set after parameter optimization as a reward signal.
}
\begin{figure*}[h]
\centering
\includegraphics[width=1\columnwidth]{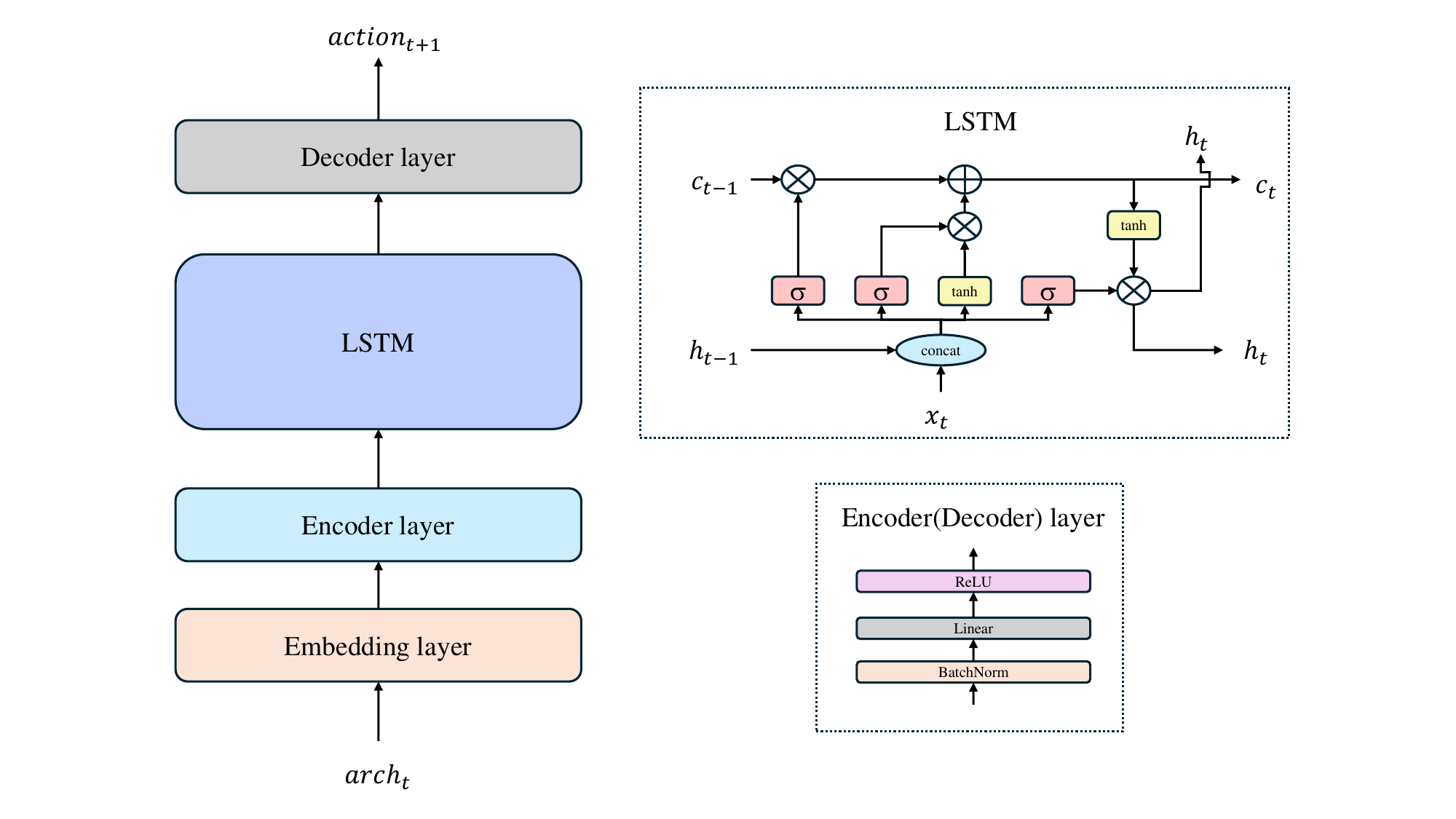}
\caption{The controller's network architecture}
\label{Fig.5}
\end{figure*}

{
The controller uses the LSTM\cite{hochreiter1997long} network to fit the policy function, which is structured as shown in Fig. 4. First to 
the input architecture sequence through the embedding layer to extract the embedded vector. Then a simple coding layer is input 
to extract a hidden layer feature with the following expression:
\begin{equation}
    out_{hidden} = ReLU(linear(BN(Embed_{input})))
\end{equation}
Where $BN$ is the regularization, $linear$ is the linear layer and $ReLU$ is the activation function. The hidden layer features 
are fed into the lstm network to get the sequence output, and then the number of quantum gates can be used by transforming the 
vectors into action dimensions through the decoding layer which has the same structure as the coding layer. Finally, the 
probability of the policy network for each action is obtained using the softmax function, and the sequence of quantum gates to 
be added to the circuit is sampled according to this probability.
}

\section{Experiments and result}

We use the pytorch \cite{paszke2019pytorch} and pennylane \cite{bergholm2018pennylane} for model development. We verify the 
performance of the model on the data classification task on the classical dataset MNIST, a handwritten image dataset, which 
is a typical entry-level machine learning task for testing various learning methods. The dataset consists of grayscale images 
of handwritten data from 0 to 9, with each image containing $28\times28$ pixels. We choose 0 and 1 for testing the binary 
classification task. The images are preprocessed and compressed to $16\times16$. The size of the original dataset is 60,000 for 
the training set, 10,000 for the test set, and 12665 and 2115 for the subsets containing only 0 and 1. We divide the 
validation set on the original training set with size 2115. Then the final training, validation, and test set sizes are 
10,550, 2115 and 2115, respectively. For the $16\times16$ pixel image, 8 quantum bits are needed to encode the input. We use 
amplitude encoding to encode the image data into qubits in the following form\:
\begin{equation}
|\psi_x\rangle=\sum_{i=1}^Nx_i|i\rangle.
\end{equation}
By amplitude encoding, the data is encoded into the amplitude of the quantum state. The data in $2^n$ dimensions can be 
encoded into $n$ qubits. We set the maximum number of quantum gates per layer to be 8, and the maximum number of layers to 
be 5. The learning rate of both circuit and controller is 0.001. The epoch of the search algorithm is 50, and the epoch of 
the training circuit to optimize the parameter subprocess is 300. The size of the randomly sampled batch of the training 
data is 25. The discount factor of the controller reward is 0.8. A simple linear layer is used for the encoding and decoding 
layers of the controller network respectively. The RNN part uses an lstm with 100 hidden units. The Adam optimizer is used 
to optimize the quantum circuit and the controller. {The final architecture obtained can reach 99.6\% accuracy on the test set, 
and just use 37 quantum gates and 11 parameters in 8 qubits.
This paper is compared with two other researches using quantum circuits for image classification, designing 
the circuit architecture by manual\cite{hur2022quantum} and searching the circuit architecture by evolutionary algorithms\cite{lu2021markovian}, and the comparison of 
the experimental results is shown in Table 1. In QCNN, 8 qubits with 30 quantum gates and 45 parameters are used. 9 qubits 
with 50 quantum gates and 106 parameters are used in Neuroevolution. In comparison, our proposed method searches the quantum 
circuit architecture using fewer qubits, while achieving good performance.}
\begin{table*}[]
    \centering
    \resizebox{\linewidth}{!}{
    \begin{tabular}{lcccccc}
    \toprule
    model & input size & encode & qubits num & gates num & parameters num & accuracy\\
    \midrule
        manual QCNN\cite{hur2022quantum} & 256 & amplitude & 8 & 30 & 45 &$98.3\%$ \\
        Neuroevolution\cite{lu2021markovian} & 256 & amplitude & 9 & 50 & 106 &$97\%$\\
        Our & 256 & amplitude & 8 & 37 & 11 &$99.6\%$\\
    \bottomrule
    \end{tabular}
    }
    \caption{MNIST dataset classification accuracy}
    \label{tab:my_label}
\end{table*}

\section{Conclusion and discussions}

We have proposed a reinforcement learning based approach in dealing with the quantum circuit structure search problem. 
Different from the previous reinforcement learning methods applied to quantum compiling, our model applies to quantum 
circuit architecture search for a given computational task. The use of layer-based search and weight sharing improves the 
computational efficiency of the search for reinforcement learning. The binary classification experiments on MNIST dataset 
show that our model can search for quantum circuits with excellent classification performance. Moreover, it has fewer 
quantum gates and variational parameters than the manually designed circuits. In addition to classical image classification 
tasks, our model can also be applied to search for circuit architectures for other computational tasks in quantum data 
processing. An interesting problem is the balance between the performance capability of the searched circuit architecture 
and  the circuit complexity, namely, to make the quantum architecture search adaptively to balance the model of reinforcement 
learning by reward design, so that the controller learns to design quantum circuits. This is also an important issue and a 
possible direction for future research.

\section*{Acknowledgments}
Thanks to Dr. Situ Haozhen from South China Agricultural University and Dr. Huang YiMing from Beijing University for their comments on this work. This work is supported by the National 
Natural Science Foundation of China under Grant Nos. 62072119, National Natural Science Foundation of China under grant 
Nos. 12075159 and 12171044, and the specific research fund of the Innovation Platform for Academicians of Hainan Province under Grant No. YSPTZX202215

\section*{Declarations}

\bmhead{Data availability}
The datasets generated during and/or analysed during the current study are available from the corresponding author on 
reasonable request.

\bmhead{Conflict of interest}
The authors have declared that they do not have any conflicts of interest.

\bibliography{reference}

\end{document}